\documentclass[showpacs, twocolumn]{revtex4-1}
\usepackage{graphicx}
\usepackage{amsmath}
\usepackage{appendix}
\usepackage{csquotes}
\usepackage{bbold}

\begin{document}

\title{Testing quantum gravity with dilute dipolar Bose gases}

\author{  Asma Tahar Taiba $^{1}$ and Abdel\^{a}ali Boudjem\^{a}a$^{2,3}$}

\affiliation{$^1$  LPTHIRM, Department of Physics, Faculty of Sciences, University of Blida 1,  P.O. Box. 270, 09000, Blida, Algeria \\
$^2$ Department of Physics, Faculty of Exact Sciences and Informatics, 
and $^3$ Laboratory of Mechanics and Energy, Hassiba Benbouali University of Chlef, P.O. Box 78, 02000, Chlef, Algeria.} 

\email {a.boudjemaa@univ-chlef.dz}

\date{\today}

\begin{abstract}

We systematically investigate the effects of quantum gravity on the ground-state properties of dilute homogeneous dipolar Bose gases using  the Hartree-Fock-Bogoliubov theory  
based on the generalized uncertainty principle. We calculate  quantum gravity corrections to the condensed fraction, the equation of state, 
the critical temperature and the superfluid fraction. 
Improved  upper  bounds on the generalized uncertainty principle parameters are found.
We compare our predictions with previous experimental and theoretical results.

\end{abstract}

	\maketitle

\section{Introduction}

Quantum theory and general relativity are the fundamental pillars of our current understanding of physics. 
They successfully describe phenomena at microscopic and macroscopic scales, respectively \cite{penrose2014}. 
Quantum gravity (QG), which aims to unify these incompatible theories, poses significant challenges in modern physics. 
One of the intriguing manifestation of QG is the so-called generalized uncertainty principle (GUP), which predicts a minimal Planck length in quantum spacetime.
The GUP is crucial in diverse QG theories and has been extensively used to  explore the Planck-scale phenomenon.
The idea of introducing a minimal length into quantum theory has a long history.
It was suggested that if one takes into account the gravitational interaction in high energy scatterings, one needs to go to higher center of mass energies (see e.g.\cite{Amati,Scard1,Adler,Lewis,Bishop} and references therein).  This is essentially a Heisenberg microscope argument \cite{Heisenberg}. 
Therefore, it turns out that the usual uncertainty relation is modified yielding  a minimal length which is one way to address the infinities that occur in certain solutions in general relativity.

The typical scales associated with QG effects is the Planck scale : $10^{16}$ TeV or $10^{-35}$ m ($\sim$14 orders of magnitude higher than the scales accessible at the LHC), 
which renders experiments to test the quantum nature of gravity prohibitive \cite{Brons}. 
However, due to rapid advances in quantum technologies tabletop tests of QG become possible suggesting that the low energy signature of QG may be detectable.
Recently, many experimental attempts aim at detecting non-classical features of the gravitational interaction have been proposed in \cite{Bose, Belen,Zych,Penrose1, Penrose2} 
using principles of quantum mechanics including entanglement, superposition, and decoherence.
In their recent work, Westphal {\it et al}. \cite{Westphal} have measured gravitational coupling between millimetre-sized masses and
Brack {\it et al}. \cite{Brack} have shown the dynamical detection of gravitational coupling between resonating beams in the hertz regime.
Most recently, Fuchs {\it et al}. \cite{Fuchs} have identified a way to measure gravity at microscopic levels using massive quantum sensors based on levitated mechanical systems,
leading to broaden our understanding of the theory of the QG.

The use of cold atoms including  Bose-Einstein condensates  (BECs) allows new ways of testing the elusive QG and its effect
due to their unprecedented degree of control and sensitivity to ultraweak forces \cite{Howl, Shir, Bris, Bris1, Dos, Das2, Hans,Jaf,Simon}.
Moreover, based on the generalized uncertainty principle (GUP) \cite{Mag,Kempf,Scar,Chang,Ali,Spre, Piko, Hus,Pedr,Feng,Shab, Gec, Scar1,Bra,Das3,Casa}, which enables for investigating QG signatures at lower and more accessible energy levels, the statistical properties of ideal Bose gases have been widely investigated 
(see, e.g., \cite{Fit, VaK,Zhang, Cast, Li, Sanj, Das} and references therein). 
The GUP approach was also employed in order to test QG in weakly interacting Bose gases \cite{Boudj}.

Motivated by the above fascinating experimental and theoretical works, we investigate in this paper low-energy QG effects using ultracold dipolar BECs
since attempts to model a full theory have not been successful until now.
Testing QG using dipolar BECs is particularly promising due to the significant role played by dipole-dipole interactions (DDIs) (see for review \cite{Pfau,Carr,Baranov}). 
Ultracold atoms with DDIs  which have the same form as the quantum gravitational interactions would be interesting  
to distinguish the QG signal from electromagnetic force \cite{Howl}, and hence offer a precise environment for testing QG effects.
The anisotropic and long-range character of DDIs make these systems especially suitable for refining the bounds on QG parameters and improving the constraints set by the GUP.

The goal is to study the effect of QG in  dilute homogeneous dipolar Bose gases using the linear and quadratic form of the GUP (LQGUP) model,
which implies a minimum measurable length and a maximum measurable momentum \cite{Ali, Das}.  
It has been shown that the LQGUP model predicts stronger QG effects compared to the purely QGUP model notably in the regime of low energies/momenta.
Furthermore, the LQGUP when applied to Bose systems, modifies the density of states \cite{Das}, and thus  the corresponding  dispersion relation, quantum and thermal fluctuations.
We calculate the QG corrections to the condensed fraction,  the critical temperature, the Lee-Huang-Yang (LHY) equation of state (EoS) and the superfluid fraction 
using the Hartree-Fock-Bogoliubov (HFB) theory based on a minimal length framework.
Our results reveal that the interplay of DDIs and QG corrections tends to improve bounds on the GUP parameters.
At higher temperatures, our results reproduce those obtained for an ideal Bose gas, as reported in \cite{Das}.

The paper is organized as follows. In Sec.~\ref{Mod}, we introduce the fundamental concepts of dipolar Bose gases. 
Section \ref{ModGup} discusses the ground-state properties of homogeneous dipolar Bose gases under the LQGUP. We compute in particular QG corrections 
to the condensed fraction, the anomalous density, the LHY corrected EoS, and to the superfluid fraction. Section \ref{Exp} delves into the experimental tests of quantum gravity.
Finally, we present our conclusions in Sec.~\ref{concl}.





\section{Theory of dipolar Bose gases} \label{Mod}

We consider a three-dimensional (3D) dilute dipolar Bose gas involving quantum and thermal fluctuations at temperature $T$. 
Assuming that the dipoles are oriented perpendicularly to the plane.
The total interaction between two atoms  of mass $m$ and dipole $d$ at $\mathbf r$ and $\mathbf r'$, can be represented as:
\begin{equation}\label{intp}
V(\mathbf r- \mathbf r')=g \delta (\mathbf r-\mathbf r')+\frac{C_{\text{dd}}} {4\pi}\frac{1-3\cos^2\theta}{|\mathbf r^3-\mathbf r^{'3}|}.
\end{equation}
The first term accounts for contact interactions component related to the $s$-wave scattering length $a$ through a coupling strength $g= (4\pi \hbar^2 a/m)$.
The second term is the DDI,  where the coupling constant $C_{\text{dd}} $ is $M_0 M^2$ for particles having a permanent magnetic dipole moment $M$ ($M_0$ is the magnetic permeability
in vacuum) and $d^2/\epsilon_0$ for particles having a permanent electric dipole $d$ ($\epsilon_0 $ is the permittivity of vacuum),
and $\theta$ is the angle between the relative position of the particles and the direction of the dipole. 
The characteristic dipole-dipole distance can be defined as $r_*=m C_{\text{dd}}/4\pi \hbar^2$ \cite {Boudj3}. 

The Hamiltonian of the system reads:
\begin{align}\label{ham}
\hat H &= \int d {\bf r} \, \hat \psi^\dagger ({\bf r}) \left(\frac{-\hbar^2 }{2m}\nabla^2+U({\bf r})-\mu\right)\hat\psi(\bf{r})  \\
&+\frac{1}{2}\int d {\bf r} \int d{\bf r'}\, \hat\psi^\dagger({\bf r}) \hat\psi^\dagger ({\bf r'}) V({\bf r-r'})\hat\psi({\bf r'}) \hat\psi(\bf{r}), \nonumber 
\end{align}
where $U({\bf r})$ is the trapping potential, $\mu$ is the chemical potential, $\hat\psi^\dagger$ and $\hat\psi$ denote, respectively the usual creation and annihilation field operators,
satisfying the usual canonical commutation relations: $[\hat\psi({\bf r}), \hat\psi^\dagger (\bf r')]=\delta ({\bf r}-{\bf r'})$.

The dynamics of dipolar BECs including the normal and anomalous fluctuations is govened by
the nonlocal generalized Gross-Pitaevskii  equation (GGPE) which can be  derived through  $i\hbar \dot{\Phi}= \partial \langle H \rangle / \partial \Phi^*$, where
$\Phi ({\bf r})=\hat\psi ({\bf r})- \hat {\bar \psi} ({\bf r})$, is the condensate wavefuncion and $\hat {\bar \psi} ({\bf r})$ is the noncondensed part of the field operator. 
This yields  \cite{Boudj4,Boudj0}:
\begin{align} \label{TDHFB}
i\hbar \dot{\Phi}& = \bigg [\frac{-\hbar^2 }{2m}\nabla^2+U({\bf r})-\mu + \delta \mu_{\text{LHY}}\\
&+ \int d{\bf r'} V({\bf r}-{\bf r'}) n ({\bf r'})  \bigg]\Phi,   \nonumber
\end{align}
where 
\begin{align} 
 \delta \mu_{\text{LHY}} ({\bf r}) \Phi({\bf r})&=\int d{\bf r'} V ({\bf r}-{\bf r'}) \bigg[\tilde n ({\bf r},{\bf r'})\Phi({\bf r'}) \label{LHYE} \\
&+\tilde m ({\bf r},{\bf r'})\Phi^*({\bf r'})\bigg],  \nonumber
\end{align}
accounts for the LHY quantum corrections to the EoS \cite{Boudj4,LHY}.
In Eq.~(\ref{LHYE}), $\tilde n ({\bf r, r'})=\langle \hat {\bar\psi}^\dagger ({\bf r}) \hat {\bar\psi} ({\bf r'}) \rangle$ and 
$\tilde m ({\bf r, r'})= \langle \hat {\bar\psi} ({\bf r}) \hat {\bar\psi} ({\bf r'}) \rangle$ are respectively,  
 the normal and anomalous correlation functions represent the dipole exchange interaction between the condensed and noncondensed atoms. 
In the local limit they reduce, respectively  to the noncondensed  $\tilde n ({\mathbf r})$ and anomalous $\tilde m ({\bf r})$ densities.
The total density is defined as: $n({\bf r})=n_c({\bf r})+\tilde n({\bf r})$, where $n_c({\bf r})=|\Phi({\bf r})|^2$ is the condensed density.

The low-lying collective excitations of dipolar BECs can be obtained within the so-called Bogoliubov-de-Gennes equations (BdGE) by considering the small oscillations of the order parameter:
$\Phi({\bf r},t)=\left[\Phi_{0} ({\bf r})+\delta\Phi({\bf r},t) \right]\exp{\left(- i\mu t/\hbar \right)}$, where
$\delta\Phi({\bf r},t)=u_{ \bf  \nu}({\bf r})\exp({-i\varepsilon_{\bf  \nu}  t/\hbar}) +v_{ \bf  \nu}({\bf r})\exp({i\varepsilon_{\bf  \nu} t/\hbar})$ are small quantum fluctuations
with $\varepsilon_{\bf  \nu}$ being the Bogoliubov excitations energy of mode $\nu$.
The quasi-particle amplitudes $ u_{ \bf  \nu}({\bf r})$ and $v_{\bf  \nu}({\bf r})$ satisfy the following nonlocal BdGEs:
\begin{align}
\varepsilon_{\bf  \nu}  u_{ {\bf \nu} } ({\bf r}) &= \hat {\cal L} u_{\bf  \nu } ({\bf r})+ \int d{\bf r'} V({\bf r}-{\bf r'}) n ({\bf r},{\bf r'}) u_{{\bf  \nu} } ({\bf r'}) \nonumber\\
&+ \int d {\bf r'}  V({\bf r}-{\bf r'}) \bar m  ({\bf r},{\bf r'}) v_{{\bf \nu} } ({\bf r'}),   \label{BdG1}  \\
-\varepsilon_{\bf \nu}  v_{ {\bf  \nu}} ({\bf r}) &= \hat {\cal L} v_{ {\bf  \nu}} ({\bf r})+ \int d{\bf r'} V({\bf r}-{\bf r'}) n ({\bf r},{\bf r'}) v_{{\bf  \nu}} ({\bf r'}) \nonumber \\
&+ \int d {\bf r'}  V({\bf r}-{\bf r'}) \bar m  ({\bf r},{\bf r'})  u_{{\bf  \nu}} ({\bf r'}), \label{BdG2}
\end{align}
where 
$\hat {\cal L} =(-\hbar^2/2m) \nabla^2+U({\bf r})-\mu + \int d{\bf r'} V({\bf r}-{\bf r'}) n ({\bf r'})$.
In the next sections, we will solve these coupled BdGE analytically for the case of a homogeneous dipolar Bose gas in the presence of QG effects.
They enable us to look at how the intriguing interplay of DDIs and QG corrections may improve bounds on the GUP parameters.

\section{Homogenenous dipolar Bose gases under the GUP} \label{ModGup}

The LQGUP taking quantum gravity effects into account is derived from the following commutation relation \cite{Ali}:
\begin{equation}\label {GUP}
[ r_i, p_j] = i\hbar  \left[ \delta_{ij}- \alpha \left( p \delta_{ij} + \frac{p_ip_j}{p}\right) + \beta \left(p^2 \delta_{ij} + 3 p_ip_j\right) \right],
\end{equation}
where  $x_i$ and $p_j$ are the position and momentum operators, respectively and $p=\sqrt{p_ip_j}$,
$\alpha= \alpha_0 l_p/\hbar=\alpha_0/(M_p c)$, and  $\beta= \beta_0 l_p^2/\hbar^2=\beta_0/(M_p c)^2$, where $\alpha_0$ and $\beta_0$ are the linear and quadratic deformation GUP parameters which are related to the Planck length, $l_p$, and the Planck mass $M_p=\sqrt{\hbar c/G}$ with $G$ being the gravitational constant and $c$ denoting the speed of light in vacuum. 
According to the authors of Ref. \cite{Ali}, the linear correction is suggested by doubly special relativity theories. 
Current experiments can set upper bounds on the GUP parameter. For instance,  
the standard model of high-energy physics implies that $\beta_0 < 10^{34}$ \cite{Das2}. 
According to the same reference \cite{Das2}, the scanning tunneling microscope delivers the best one $\beta_0 < 10^{21}$ \cite{Das2}. 
Other upper bounds have been provided by different approaches, namely the Lamb shift and Landau levels \cite{Das2}, optical systems \cite{Bra},  
the light deflection and perihelion precession \cite{Scard}, cold atoms \cite{Gao}, and gravitational systems \cite{Feng1,Nev,DD}.
However, in the present work we will address QG effects for arbitrary $\beta_0$.
For $\alpha=\beta=0$, one can reproduce the standard Heisenberg uncertainty principle. 

One should stress that Eq.~(\ref{GUP}) implies that standard operator for momentum  cannot be used, $p_i \neq -i\hbar \nabla$. 
However, we introduce a set of canonical operators $x_{0i}$ and $p_{0i}$, which satisfy a standard commutation relation $[r_{0i}, p_{0i}]= i\hbar \delta_{ij} $ \cite{Ali}.
Doing so, we can write 
\begin{equation}\label {GUP1}
r_i= r_{0i},   \;\;\;\;\;\;\;  p_i= p_{0i} (1-\alpha p_0+ 2\beta p_0^2),
\end{equation}
where $p=\sqrt{p_{0i} p_{0i}}$.\\
Furthermore, according to Eq.~(\ref{GUP}), the deformed density of states, $g(E)$, can be given by \cite{Das}
	\begin{align}\label {GUP2}
	\sum_{\mathbf n}& =\frac{V}{(2\pi \hbar)^3} \int_{0}^{\infty} d^3p \\
&= \frac{1}{2}\int_{0}^{\pi} \sin{\theta}d\theta \int_{0}^{\infty} g(E)\,dE, \nonumber
	\end{align}
where 
\begin{equation} \label{DoS}
g(E)=\frac{(2m)^{3/2}}{4\pi^2 \hbar^3} E^{1/2}(1+16\alpha\sqrt{m}E^{1/2}-25\beta m E),
\end{equation} 
with $E=p^2/2m$ being the energy of free particle \cite{Das}.
Equations (\ref{GUP1}) and (\ref{GUP2}) show that the LQGUP can alter not only the excitations 
energy but also the statistical and the thermodynamic properties of a weakly interacting dipolar Bose gas. 

For concreteness, let us consider a homogeneous 3D Bose gas ($U(\mathbf r)=0$) with DDI enclosed in a volume $V$. 
These unifrom systems constitute a prototype for the experimentally relevant trapped ultracold gases and often lead to the correct physical intuition with respect to their properties.
In such a case, the densities $n_c$, $\tilde n$, and $\tilde m$ are constant and the momentum-space interaction is independent of the magnitude of $\mathbf p$ 
and instead depends only on its direction.

From now on we will replace $p_0$ by $p$ for simplicity.
Taking into account effects of QG  implied by the LQGUP given in Eq.~(\ref{GUP1}), the GGPE (\ref{TDHFB}) turns out to be given as:
\begin{equation} \label{TDHFB2}
i\hbar \dot{\Phi} = \bigg [{\cal E}_{\mathbf p}+ V(|\mathbf p|=0) n + \delta \mu_{\text{LHY}} (\mathbf p) -\mu \bigg]\Phi,  
\end{equation}
where ${\cal E}_p= (p^2 /2m) \left(1- 2\alpha  p+5\beta p^2 \right)$, and 
\begin{equation}\label {DDFour}
V ({\mathbf p}) =g[1+\epsilon_{\text{dd}}(3\cos_{\mathbf p}^2\theta-1)],
\end{equation}
is the Fourier transform of the interaction potential (\ref{intp}), 
$\epsilon_{\text{dd}}=C_{\text{dd}}/3g$ is the dimensionless relative strength describing the interplay between the DDI and contact interaction, and 
$\theta$ being the angle between the vector $\mathbf p$ and the polarization direction.
Note that the relation (\ref{DDFour}) is valid only in the  ultracold limit where the particle momenta satisfy the inequality $p r_*/\hbar \ll 1$ \cite{Boudj}. 

The chemical potential is obtained via Eq.~(\ref{TDHFB2}) \cite{Boudj0}
\begin{align} \label{chim0}
\mu&=V(|\mathbf p|=0) n + \delta \mu_{\text{LHY}} (\mathbf p) \\
&=V(|\mathbf p|=0) n + \frac{1}{V}\sum\limits_{\bf p \neq 0} V(\mathbf p) \big(\tilde n_{p} +\tilde m_{p} \big), \nonumber
\end{align}
where $\tilde n_p$ and  $\tilde m_p$  stand for the normal and anomalous distributions (see below).

\subsection{Excitations energy}

\begin{figure}
\centering 
\includegraphics[scale=0.8, angle=0] {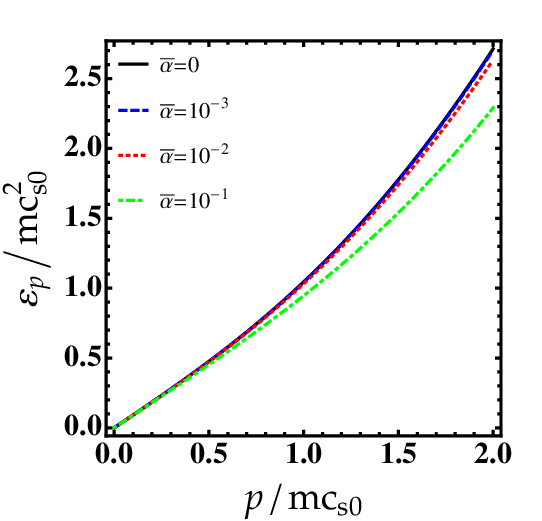}
\caption { Bogoliubov spectrum for different values of the GUP parameters, $\overline\alpha$, with $\theta=\pi/2$, $\epsilon_{\text{dd}}=0.16$.
Here we set $\overline \alpha= \alpha\, m c_{s0}= \alpha_0 \left(m c_{s0}/ M_p c \right)$ and $\beta_0=\alpha_0^2$ \cite{Das}.}
\label{BBS}
\end{figure}

The weakly interacting regime suggests $\tilde n/n \ll 1$ and $\tilde m/n \ll 1$ \cite{Boudj1,Boudj2}.
Then, the Bogoliubov excitations energy  can be obtained  via the BdGEs (\ref{BdG1}) and (\ref{BdG2}), where the  Bogoliubov quasiparticle amplitudes
take the form $u_p,v_p =(\sqrt{\varepsilon_p /{\cal E}_p} \pm \sqrt{{\cal E}_p/\varepsilon_p})/2$ \cite{Bog}.
After some algebra we find the QG corrected Bogoliubov dispersion relation:
\begin{equation}\label{BogR} 
\varepsilon_p= \sqrt{ {\cal E}_p^2 + 2 V(\mathbf p) {\cal E}_p}.
\end{equation}
For small momenta $p\rightarrow 0$, the Bogoliubov dispersion relation is phonon-like $\varepsilon_p= c_s (\theta) p$ (quanta of sound waves), 
where  $c_s(\theta)= c_{s0}\sqrt{ 1+\epsilon_{\text{dd}}(3\cos^2\theta-1)}$ is the sound velocity which is anisotropic owing to the DDI, with $c_{s0}=\sqrt{ g n/m}$ being the sound velocity of a nondipolar BEC. 
In the high momenta limit $p\rightarrow \infty$, the excitations spectrum (\ref{BogR}) reduces to  $\varepsilon_p={\cal E}_P$. 
For $\alpha=\beta=0$, one recovers the standard Bogoliubov energy \cite{Bog}.
It is important to note that, for $\epsilon_{\text{dd}}>1$, the spectrum (\ref{BogR}) becomes imaginary giving rise to the collapse of homogeneous dipolar BEC with dominant DDI.
Effects of QG on the Bogoliubov excitations spectrum are singinifcant notably for large momenta, $p/mc_{s0} \geq 1$, as shown in Fig.~\ref{BBS}.

\subsection{Condensed fraction and transition temperature}

In the realm of the Bogoliubov theory, $\tilde n_p$ and $\tilde m_p$ are defined as:
\begin{equation}\label{Ndiso} 
\tilde n_p=v_p^2+ (u_p^2+v_p^2)N_p,
\end{equation}
and 
\begin{equation}\label{Mdiso} 
 \tilde m_p=u_p v_p (2N_p+1),
\end{equation}
where $N_p=[\exp(\varepsilon_p/T)-1]^{-1}$  are occupation numbers for the excitations. 
Therefore, explicit formulas for the noncondensed and the anomalous densities can be obtained using  the definitions $\tilde{n}=V^{-1}\sum_{\mathbf  p} \tilde n_p$ and 
$\tilde{m}=-V^{-1}\sum_{\mathbf  p} \tilde m_p$, and the QG corrected density of states in Eq.~(\ref{DoS}):
\begin{align}\label {nor}
\tilde{n}&=\frac{1}{4}\int_{0}^{\pi} \sin{\theta}d\theta \int_{0}^{\infty} g(E)\,dE \left[\frac{{\cal E}_E+ m c_s^2(\theta)} {\varepsilon_E}\right]\\
& \times \left[\coth\left(\frac{\varepsilon_E}{2T}\right)-1\right], \nonumber
\end{align}
and
\begin{equation}\label {anom}
\tilde{m}=- \frac{1}{4}\int_{0}^{\pi} \sin{\theta}d\theta \int_{0}^{\infty} g(E)\,dE \frac{ m c_s^2 (\theta)} {\varepsilon_E}\coth\left(\frac{\varepsilon_E}{2T}\right).
\end{equation}
Leading terms in Eqs.(\ref{nor}) and (\ref {anom}) are the zero-temperature contribution to the noncondensed $\tilde{n}_0$ and anomalous $\tilde{m}_0$ densities, respectively.
Subleading terms represent the contribution of the so-called  thermal fluctuations and we denote them as  $\tilde{n}_T$ and $\tilde{m}_T$, respectively. 
Equations (\ref{nor})  and (\ref{anom}) allow us to determine in a very useful way the condensed fraction and the critical temperature of Bose quantum liquids (see below).

\begin{figure}
\centering 
\includegraphics[scale=0.45, angle=0] {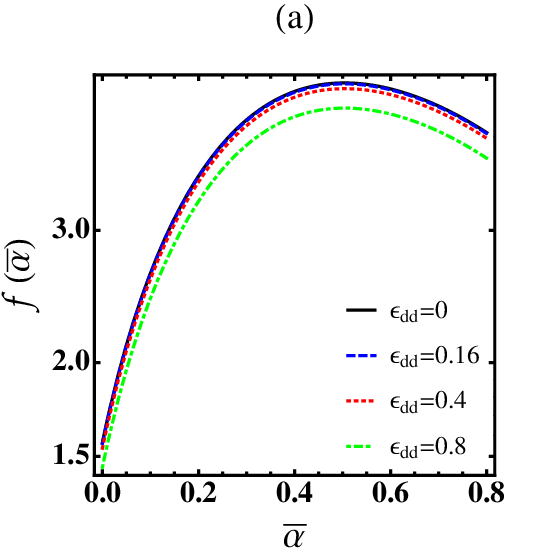}
\includegraphics[scale=0.45, angle=0] {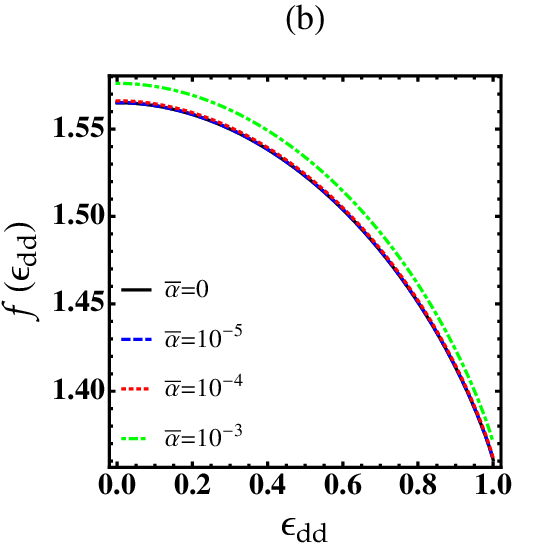}
\caption { Deformation function $f (\overline\alpha, \epsilon_{\text{dd}})$ which governs the dependence of the condensate depletion as a function of the deformation parameter $\alpha$ in units of 
$m c_{s0}$.}
\label{DpS}
\end{figure}

\begin{figure}
\centering 
\includegraphics[scale=0.45, angle=0] {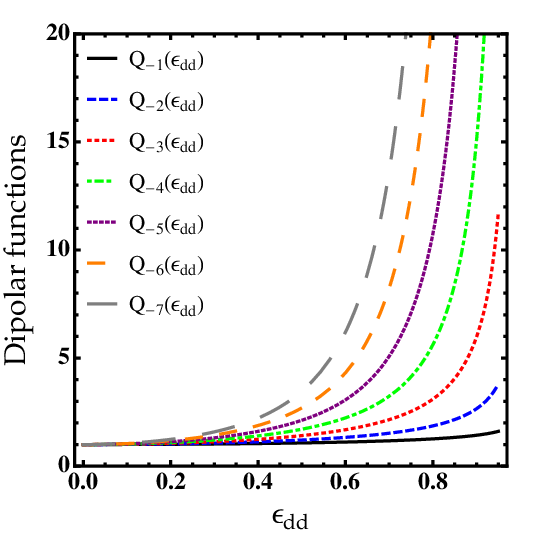}
\includegraphics[scale=0.45, angle=0] {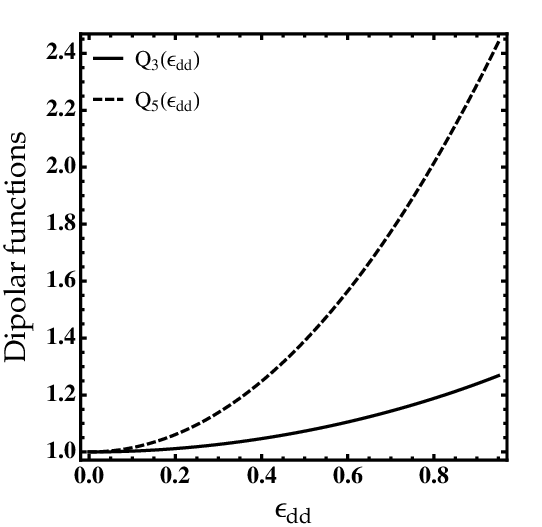}
\caption { Dipolar functions ${\cal Q}_j$ vs. the dipolar interaction parameter $\epsilon_{\text{dd}}$.}
\label{DF}
\end{figure}

At zero temperature, integral (\ref{nor}) gives for the quantum depletion:  $\tilde n_0= (1/3\pi^2) (m c_{s0}/ \hbar)^3 f (\overline\alpha, \epsilon_{\text{dd}})$, where
\begin{widetext}
\begin{align}\label {norT0}
f (\overline\alpha, \epsilon_{\text{dd}})&= \frac{3} {4 \sqrt{2}}  \int_0^{\pi} \sin \theta  d\theta \int_0^{\infty} dx \sqrt{x} \left(-25 \overline\beta x^2+16 \sqrt{x} \overline\alpha+1\right) \\
&\times \left(\frac{ (x^2/2) \left(5 \overline\beta x^2 -2 \overline\alpha x+1\right)+\epsilon_{\text{dd}}  \left(3\cos ^2 \theta-1\right)+1}
{\sqrt{x^2 \left(5  \overline\beta x^2-2 \overline\alpha x+1\right) 
\left(\epsilon_{\text{dd}}  \left(3 \cos^2 \theta-1\right)+1\right)+( x^4/4)\left(5 \overline\beta x^2 -2 \overline\alpha x +1\right)^2}}-1\right), \nonumber
\end{align}
\end{widetext}
where $\overline \alpha=\alpha\, m c_{s0}= \alpha_0 \left(m c_{s0}/ M_p c \right)$, and $\overline \beta= \beta\, (m c_{s0})^2= \beta_0 \left(m c_{s0}/ M_p c \right)^2$.
The result of the numerical integration of Eq.~(\ref{norT0}) is shown in  Fig.~\ref{DpS}.
We see that the function $f$ increases for $\overline\alpha \lesssim 0.5$, reaches its maximum at  $\overline\alpha \sim 0.5$, then it decreases for larger $\overline\alpha$ 
regardless of the value of the DDI strength, $\epsilon_{\text{dd}}$ (see Fig.~\ref{DpS}.a).
This clearly reveals that the QG effects may significantly modify the quantum depletion and the condensed fraction of the condensate.   
Surprisingly, the deformation function $f $ is lowering with $\epsilon_{\text{dd}}$ which is in contrast to dipolar BEC without GUP (see Fig.~\ref{DpS}.b). 
For $\overline\beta=\overline\alpha=0$, we reproduce the results of  the depletion for a dipolar BEC without GUP, $\tilde n_0 =(1/3\pi^2) (m c_{s0}/ \hbar)^3 Q_3 (\epsilon_{\text{dd}})$ 
\cite{Axel, Boudj1,Boudj2}, 
where the DDI contribution is described by the function ${\cal Q}_3(\epsilon_{\text{dd}})$ (see Fig.~\ref{DF} (right panel)), which is special case $j=3$ of
 ${\cal Q}_j(\epsilon_{\text{dd}})=(1-\epsilon_{\text{dd}})^{j/2} {}_2\!F_1\left(-\frac{j}{2},\frac{1}{2};\frac{3}{2};\frac{3\epsilon_{\text{dd}}}{\epsilon_{\text{dd}}-1}\right)$, where ${}_2\!F_1$ 
is the hypergeometric function. Note that functions ${\cal Q}_j(\epsilon_{\text{dd}})$ attain their maximal values for $\epsilon_{\text{dd}}\approx 1$ and become imaginary for $\epsilon_{\text{dd}}>1$.\\
For vanishing quadratic QG correction ($\overline\alpha=0$) and for $\epsilon_{\text{dd}}=0$,  one can expect that the depletion (\ref{norT0}) reduces to our recent result \cite{Boudj3}.

At low temperatures $T \ll m c_{s0}^2$, the main contribution to integral (\ref{nor}) comes from the low energy branch where
${\cal E} \approx E$ and  $\varepsilon \approx c_E(\theta) \sqrt{2mE}$. 
A straightforward calculation gives for the condensed fraction, $n_c/n=1- \tilde n/n$:
\begin{align} 
\frac{n_c}{n}&=1-\left(\frac{T}{T_c^{0}} \right)^2\bigg[\frac{\pi ^2  {\cal Q}_{-1} (\epsilon_{\text{dd}})}{3 \zeta (3/2)^{4/3}} \left(\xi n^{1/3}\right)   \label{confra} \\
&+\overline\alpha  \frac{32  \pi \zeta (3)  {\cal Q}_{-2}(\epsilon_{\text{dd}}) }{\sqrt{2} \zeta (3/2)^{2/3}} \left(\xi n^{1/3}\right) ^3 \left(\frac{T}{T_c^{0}} \right) \nonumber\\
&-\overline\beta \frac{10 \pi ^6 {\cal Q}_{-3}(\epsilon_{\text{dd}})}{3 \zeta (3/2)^{8/3}}  \left(\xi n^{1/3}\right) ^5 \left(\frac{T}{T_c^{0}} \right)^2 \bigg],\nonumber
\end{align}
where $\xi = \hbar/(mc_{s0})$ is the healing length and $T_c^0= (2 \pi \hbar^2/\zeta (3/2)^{2/3}m) n^{2/3}$ is the ideal gas transition temperature.
In Eq.~(\ref{confra}) we utilized the identity $\int_0^{\infty}  x^j dx /(e^{x}-1) =\Gamma (j+1) \zeta (j+1)$, where $\Gamma (x)$ is the gamma function and 
$\zeta (x)$ is the Riemann zeta function.
For $\overline\beta=\overline\alpha=0$, one recovers the results of the depletion for a dipolar BEC without GUP, $\tilde n_T =(m T^2/12 \hbar^3c_{s0}) Q_{-1} (\epsilon_{\text{dd}})$ \cite{Boudj1,Boudj2}.
The behavior of the functions $Q_{-1} (\epsilon_{\text{dd}})$, $Q_{-2} (\epsilon_{\text{dd}})$, and $Q_{-3} (\epsilon_{\text{dd}})$ is displayed in Fig.~\ref{DF}.
Equation (\ref{confra}) shows also that the QG corrections (second and last terms in r.h.s) increase with increasing  the density $n$ and temperature, $T/T_c^0$, 
and with decreasing boson mass $m$.

\begin{figure}
\centering 
\includegraphics[scale=0.45, angle=0] {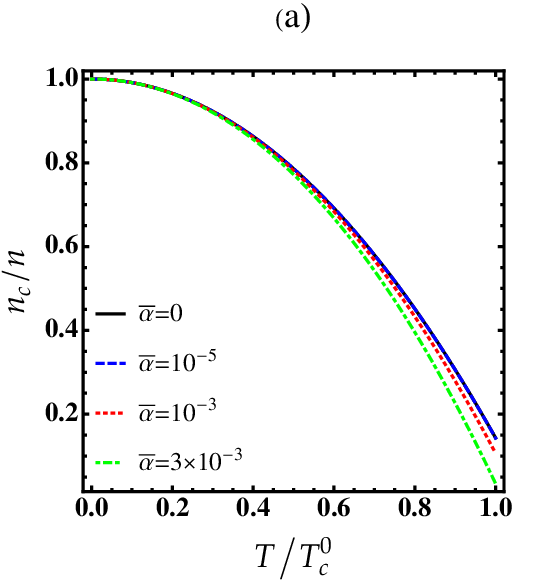}
\includegraphics[scale=0.45, angle=0] {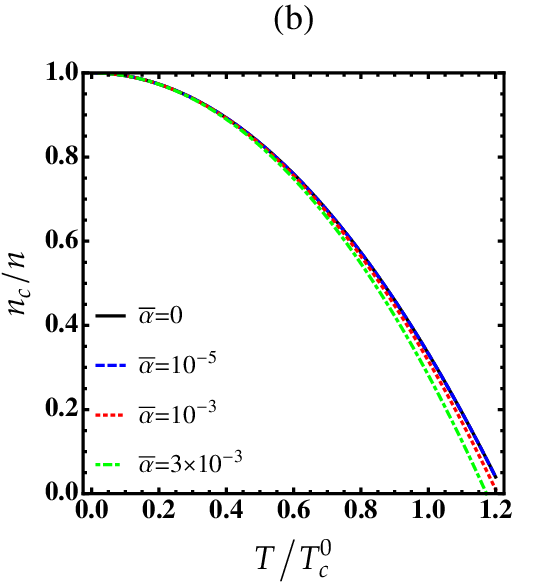}
\includegraphics[scale=0.45, angle=0] {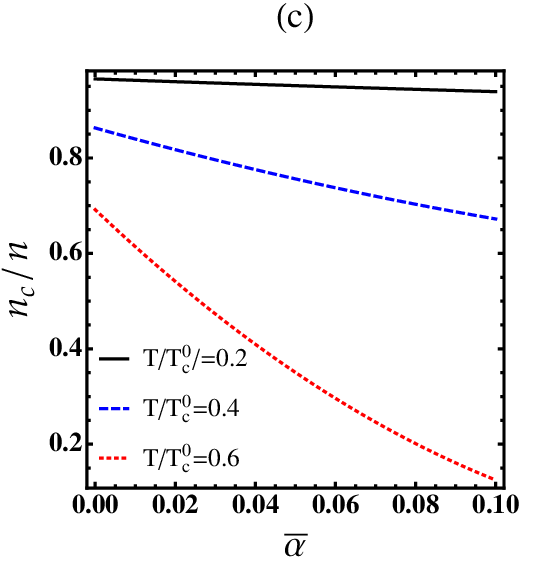}
\includegraphics[scale=0.46, angle=0] {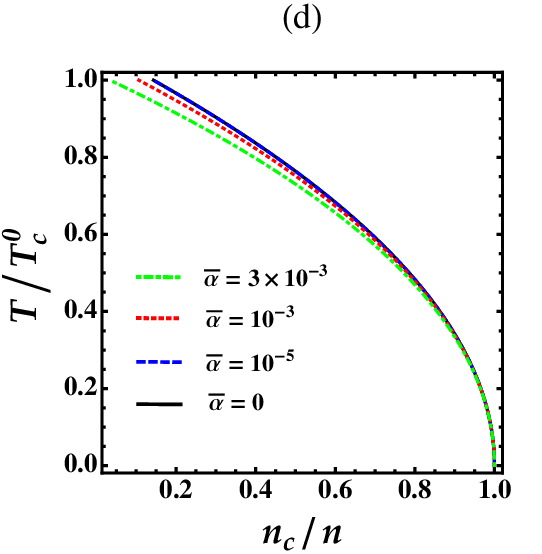}
\caption { (a) Condensed fraction of Cr BEC, $n_c/n$, as a function of the reduced temperature, $T/T_c^0$, for different values of $\overline\alpha$ with $\epsilon_{\text{dd}}=0.16$
and  $\xi n^{1/3}= 0.93$ \cite{Pfau}. 
(b) The same as Fig.~\ref{CF}.a but for Er BEC with $\epsilon_{\text{dd}}=0.38$ and $\xi n^{1/3}= 0.7$ \cite{erbium}.
(c) Condensed fraction as a function of the GUP parameter, $\overline\alpha$, for different values of reduced temperature, $T/T_c^0$, with $\epsilon_{\text{dd}}=0.16$
and  $\xi n^{1/3}= 0.93$. 
(d) Reduced temperature $T/T_c^0$ as a function of the  condensed fraction, $n_c/n$, for different values of $\overline\alpha$ with  $\xi n^{1/3}= 0.93$ and $\epsilon_{\text{dd}}=0.16$.}
\label{CF}
\end{figure}

To illustrate our results, we consider two cases namely: ${}^{52}$Cr atoms  with $\epsilon_{\text{dd}}=0.16$ \cite{Pfau} and ${}^{168}$Er atoms with $\epsilon_{\text{dd}}=0.38$ \cite{erbium}.

Figures \ref{CF} (a) and (b) depict that the condensed fraction $n_c/n$ decreases with the reduced temperature, $T/T_c^0$, regardless of the 
DDI relative strength, $\epsilon_{\text{dd}}$.
It decays also with the GUP parameter $\overline\alpha$ notably for large $T$ as shown in Fig.~\ref{CF} (c) indicating that
QG effects may reduce the condensed density and thus enhance the thermal cloud. 
For instance,  at $T=0.2 T_c^0$, one has  $n_c/n \simeq  98\%$  for $\overline\alpha =0$ then it reduces to $n_c/n \simeq  94\%$  for $\overline\alpha =0.1$.
Whereas at $T=0.6 T_c^0$, the condensed fraction is about $n_c/n \simeq  70 \%$ while it decays to $n_c/n \simeq  10 \%$  for $\overline\alpha =0.1$.
In  Fig.~\ref{CF} (d) we plot the reduced temperature $T/T_c^0$ as a function of the  condensed fraction, $n_c/n$, for different values of $\overline\alpha$.
We see that the GQ  corrections lower $T/T_c^0$ only for relatively small $n_c/n$. For large $n_c/n$, the reduced temperature is almost insensitive to the GQ effects.
This downshift is a clear indication of the significance of the GQ effects in particular at higher temperature when thermal fraction becomes important.


However, at higher temperatures, $T \gg m c_{s0}^2$, where the main contribution to integral (\ref{nor}) comes from the high energy branch, our results coincide with those obtained 
in \cite{Ali} for ideal Bose gases.

\subsection{Equation of state}

Corrections to the chemical potential due to the LHY quantum fluctuations can be derived by combining  Eqs.~ (\ref{Ndiso}), (\ref{Mdiso}) and (\ref{chim0}).
This yields:
\begin{align} \label{chim01}
\mu_{\text{LHY}}&=\frac{1}{4}\int_{0}^{\pi} \sin{\theta}d\theta \int_{0}^{\infty} g(E)\,dE \,\tilde V(E) \\
 &\times\left[\frac{ {\cal E}_E} {\varepsilon_{E}} \coth\left(\frac{\varepsilon_E}{2T}\right)-1\right]. \nonumber
\end{align}
This equation is appealing since it enables us to calculate the LHY corrections to all thermodynamic quantities.
At low temperatures, the LHY corrected EoS can be written as:
\begin{align} \label{chim11}
\mu_{\text{LHY}}&=\frac{\pi^2gT^4}{60 m \hbar^3 c_{s0}^5} 
\bigg[\frac{ {\cal Q}_{-3} (\epsilon_{\text{dd}})} {8} +\overline\alpha  \frac{ 720 \zeta (5) {\cal Q}_{-4} (\epsilon_{\text{dd}})  }{\sqrt{2} }  \left(\frac{T} {c_{s0} } \right)  \nonumber\\
&-\overline\beta \frac{125 \pi ^6 {\cal Q}_{-5}(\epsilon_{\text{dd}}) } {42}  \left(\frac{T} {c_{s0}} \right)^2 \bigg]. 
\end{align}
In terms of the reduced temperature, $\mu_{\text{LHY}}$ reads
\begin{align} \label{chim12}
\frac{\mu_{\text{LHY}}} {ng}&=\left(\frac{T}{T_c^0}\right)^4
\bigg[\frac{\pi^6 {\cal Q}_{-3} (\epsilon_{\text{dd}})} {30 \zeta (3/2)^{8/3}} (\xi n^{1/3})^5 \\
&+\overline\alpha  \frac{ 384 \pi^7 \zeta (5) {\cal Q}_{-4} (\epsilon_{\text{dd}})  }{\sqrt{2} \zeta (3/2)^{10/3}}  (\xi n^{1/3})^7 \left(\frac{T}{T_c^0}\right) \nonumber\\
&-\overline\beta \frac{200 \pi ^{10} {\cal Q}_{-5}(\epsilon_{\text{dd}}) } {63\zeta (3/2)^{4}}  (\xi n^{1/3})^9 \left(\frac{T}{T_c^0}\right)^2 \bigg]. \nonumber
\end{align}
Again for $\overline\beta=\overline\alpha=0$, $\mu_{\text{LHY}}$  of Eq.~(\ref{chim12}) simplifies to that for a dipolar BEC without GUP,  
$g \pi^2T^4 {\cal Q}_{-3} (\epsilon_{\text{dd}}) / (60 m \hbar^3 c_{s0}^5)$ \cite{Boudj1,Boudj2}.
The behavior of the functions $Q_{-5} (\epsilon_{\text{dd}})$, and $Q_{-7} (\epsilon_{\text{dd}})$ is shown in Fig.~\ref{DF}.

 \begin{figure}
\centering 
\includegraphics[scale=0.46, angle=0] {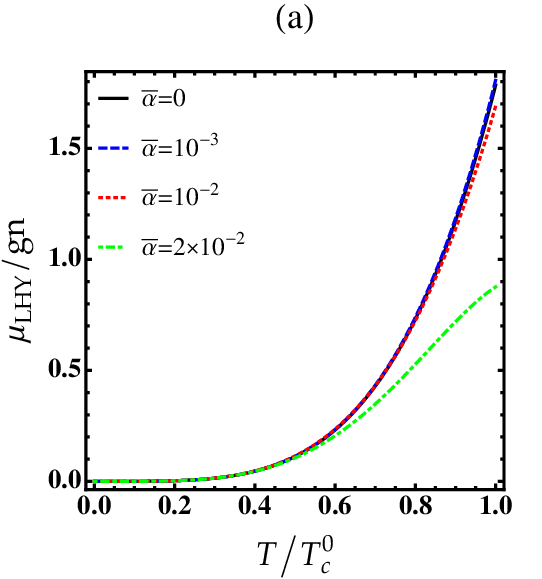}
\includegraphics[scale=0.46, angle=0] {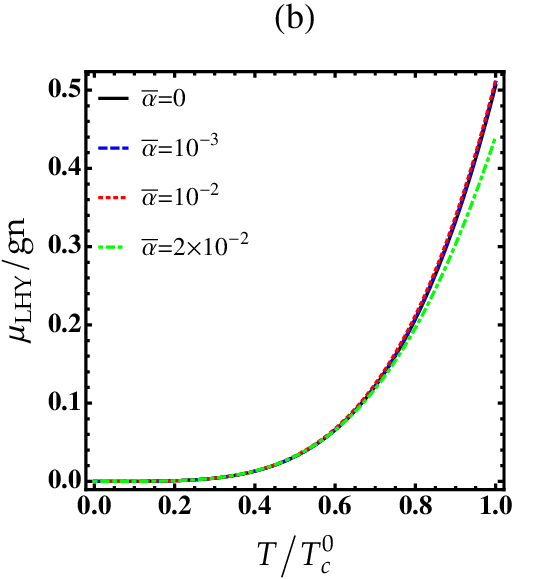}
\caption { (a) Thermal contribution to the LHY corrected EoS of Cr BEC, $\mu_{\text{LHY}}/(ng)$, as a function of the reduced temperature, $T/T_c^0$, for different values of $\overline\alpha$ 
with $\epsilon_{\text{dd}}=0.16$ and $\xi n^{1/3}= 0.93$ \cite{Pfau}. 
(b) The same as Fig.~(\ref{LHY}.a)  but for Er BEC with $\epsilon_{\text{dd}}=0.38$ and $\xi n^{1/3}= 0.7$ \cite{erbium}.}
\label{LHY}
\end{figure}

Figure \ref{LHY} shows that for $\overline\alpha  \gtrsim 2 \times 10^{-2}$, the thermal contribution to the LHY corrected EoS, $\mu_{\text{LHY}}/(ng)$, deviates from that 
of a dipolar BEC without GUP at higher temperatures owing to QG effects. Remarkably, this discrepancy becomes pronounced for small $\epsilon_{\text{dd}}$ (see Fig.~\ref{LHY} (a)).

\subsection{Superfluidity}

\begin{figure}
\centering 
\includegraphics[scale=0.45, angle=0] {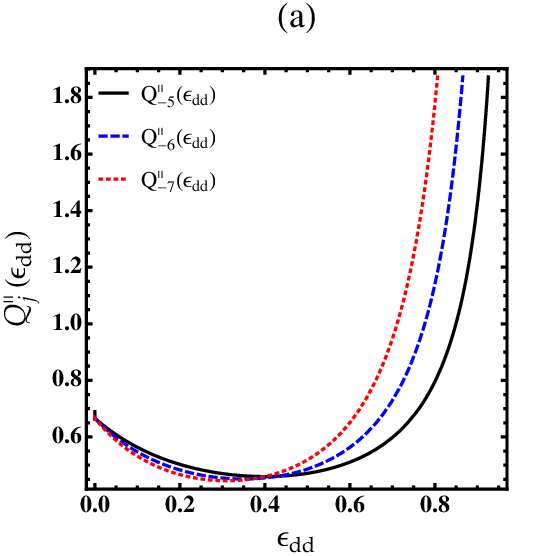}
\includegraphics[scale=0.45, angle=0] {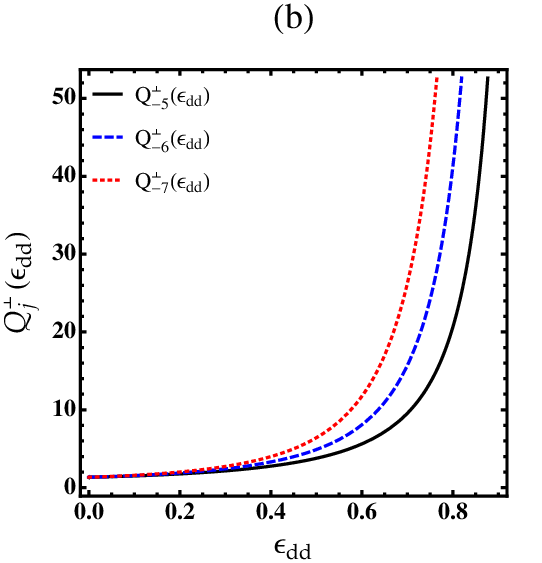}
\caption {Parallel ${\cal Q}_j^{\parallel} (\epsilon_{\text{dd}})$ (a) and perpendicular ${\cal Q}_{j}^{\perp}(\epsilon_{\text{dd}})$ (b) dipolar functions  and vs. the dipolar interaction parameter $\epsilon_{\text{dd}}$.}
\label{DFS}
\end{figure}

Now we focus on  the behavior of the superfluid fraction of a dipolar BEC under GUP.
In a 3D dipolar BEC the superfluid density $n_s$ is a tensor quantity with components $n_s^{ij}$ due to the peculiar anisotropy property of the DDI \cite{Axel3,Boudj5}.
This means that $n_s$ depends on the direction of the superfluid motion with respect to the orientation of the dipoles.
It can be found by applying a Galilean boost with the total momentum of the moving system ${\bf P}=m V(n {\bf v_s}+n_n {\bf v_n})$, where 
${\bf v_s}$ denotes the superfluid velocity and ${\bf v_n}={\bf u}-{\bf v_s}$ is normal fluid  velocity with ${\bf u}$ being a boost velocity \cite{Axel3,Boudj5}. 
Keeping only linear term in $P$, we get
\begin{equation} \label{sup}
 \frac{n_s^{ij}}{n}= \delta_{ij}- \frac{1}{Tnm}\int \frac{d^3 p}{(2 \pi\hbar)^3} \left[\frac{ p_i p_j}{4\, \text {sinh}^2 (\varepsilon_k/2T)}\right]. 
\end{equation}
Applying the QG corrected density of states in Eq.~(\ref{DoS}),  the superfluid fraction (\ref{sup}) turns out to be given: 
\begin{equation} \label{sup1}
 \frac{n_s^{ij}}{n}= \delta_{ij}- \frac{1}{Tnm} \int_{0}^{\pi} \sin{\theta}d\theta \int_{0}^{\infty} g(E)\,dE \left[\frac{\sqrt{E_i\, E_j}}{4\, \text {sinh}^2 (\varepsilon_k/2T)}\right]. 
\end{equation}
At low temperatures $T\ll n g$, the parallel direction of the superfluid fraction reads
\begin{widetext}
\begin{align} \label{supflui1}
\frac{n_s^{\parallel}}{n} =1-\left(\frac{T}{T_c^0}\right)^4
\bigg[\frac{\pi^7 {\cal Q}_{-5}^{\parallel} (\epsilon_{\text{dd}})} {30 \zeta (3/2)^{8/3}} (\xi n^{1/3})^5
+\overline\alpha  \frac{ 480 \pi^4 \zeta (5) {\cal Q}_{-6} ^{\parallel}(\epsilon_{\text{dd}})  }{\sqrt{2} \zeta (3/2)^{10/3}}  (\xi n^{1/3})^6 \left(\frac{T}{T_c^0}\right) 
-\overline\beta \frac{25 \pi ^{11} {\cal Q}_{-7}^{\parallel}(\epsilon_{\text{dd}}) } {8\zeta (3/2)^{4}}  (\xi n^{1/3})^9 \left(\frac{T}{T_c^0}\right)^2 \bigg]. 
\end{align}
where the functions
\begin{align}
{\cal Q}_{j}^{\parallel} (\epsilon_{\text{dd}})&= \frac{2 (1-\epsilon_{\text{dd}})^{j/2}}{9 (j+1) (j+3) \epsilon_{\text{dd}} ^2}  \left[(\epsilon_{\text{dd}} -1)^2 
\, _2F_1\left(-\frac{1}{2},-\frac{j}{2};\frac{1}{2}; \frac{3 \epsilon_{\text{dd}} }{\epsilon_{\text{dd}} -1}\right)+(2\epsilon_{\text{dd}} +1) ((3 j+4) \epsilon_{\text{dd}} -1) \left(1-\frac{3 \epsilon_{\text{dd}}}{\epsilon_{\text{dd}} -1}\right)^{j/2}\right],\nonumber
\end{align}
behave as ${\cal Q}_j^{\parallel} (\epsilon_{\text{dd}}=0)=2/3$ and imaginary for $\epsilon_{\text{dd}}>1$ (see also Fig.~\ref{DFS}.a). \\
In the perpendicular direction, one has
\begin{align} \label{supflui2}
\frac{n_s^{\perp}}{n} =1-\left(\frac{T}{T_c^0}\right)^4
\bigg[\frac{\pi^7 {\cal Q}_{-5}^{\perp} (\epsilon_{\text{dd}})} {30 \zeta (3/2)^{8/3}} (\xi n^{1/3})^5
+\overline\alpha  \frac{ 480 \pi^4 \zeta (5) {\cal Q}_{-6} ^{\perp}(\epsilon_{\text{dd}})  }{\sqrt{2} \zeta (3/2)^{10/3}}  (\xi n^{1/3})^6 \left(\frac{T}{T_c^0}\right) 
-\overline\beta \frac{25 \pi ^{11} {\cal Q}_{-7}^{\perp}(\epsilon_{\text{dd}}) } {8\zeta (3/2)^{4}}  (\xi n^{1/3})^9 \left(\frac{T}{T_c^0}\right)^2 \bigg]. 
\end{align}
where ${\cal Q}_{j}^{\perp}(\epsilon_{\text{dd}})= {\cal Q}_{j}(\epsilon_{\text{dd}})-{\cal Q}_{j}^{\parallel} (\epsilon_{\text{dd}})$ which increases monotonically with $\epsilon_{\text{dd}}$ and 
becomes complex for $\epsilon_{\text{dd}}>1$ (see Fig.~\ref{DFS}.b).
For $\overline\beta=\overline\alpha=0$, the superfluid fraction in both directions reduces to that obtained for a dipolar BEC without GUP \cite{Axel3,Boudj5}.
\end{widetext}

\begin{figure}
\centering 
\includegraphics[scale=0.45, angle=0] {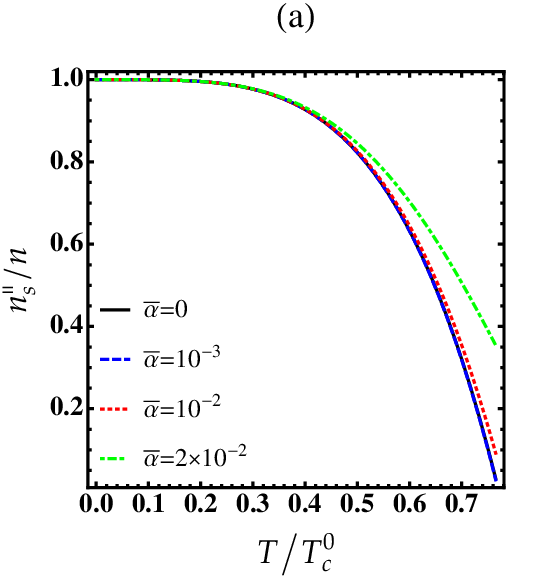}
\includegraphics[scale=0.45, angle=0] {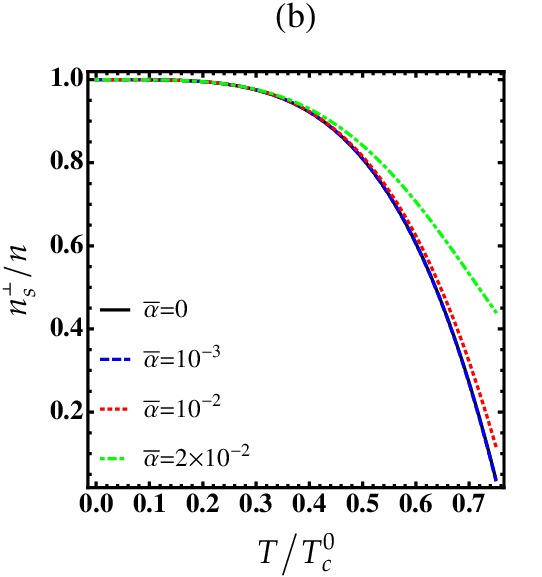}
\includegraphics[scale=0.45, angle=0] {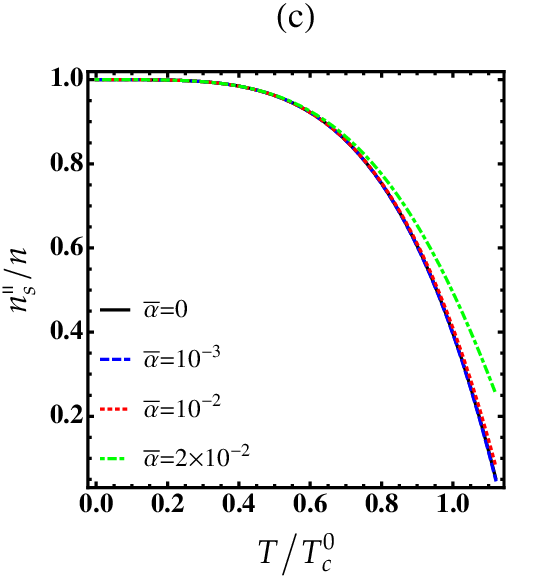}
\includegraphics[scale=0.45, angle=0] {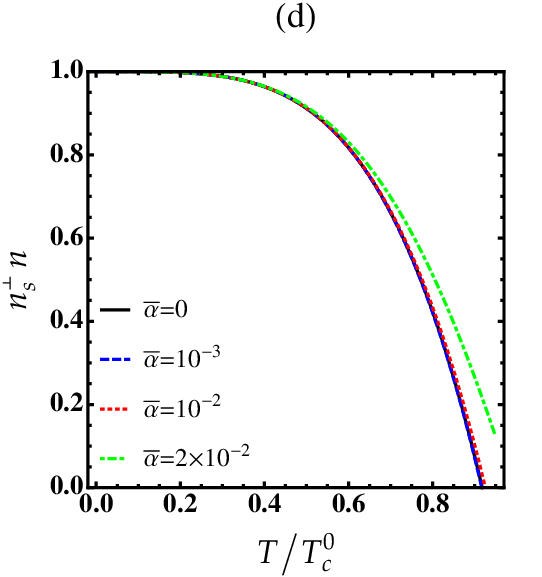}
\caption { Superfluid fractions $n_s^{\parallel}/n$ (a) and $n_s^{\perp}/n$ (b) of Cr BEC, as a function of the reduced temperature, $T/T_c^0$, for different values of $\overline\alpha$ 
with  $\epsilon_{\text{dd}}=0.16$ and $\xi n^{1/3}= 0.93$ \cite{Pfau}. 
(c) -(d) The same as Figs.~\ref{CF}.(a) and (b)  but for Er BEC with $\epsilon_{\text{dd}}=0.38$ and $\xi n^{1/3}= 0.7$ \cite{erbium}.}
\label{CFS}
\end{figure}

Figure \ref{CFS} depicts that the superfluid fraction is decreasing with the reduced temperature in both directions, regardless of the value of the GUP parameter and of DDI strength.
A direct comparison between both components shows that  $n_s^{\parallel}$ and  $n_s^{\perp}$ coincide for $\overline\alpha=0$ and for $\epsilon_{dd}=0$
and hence, well reproduce the standard two-body contact interaction result.
We see also that  for relatively high temperatures  $T \gtrsim 0.45 T_c^0$ and for $\epsilon_{\text{dd}}=0.16$, $n_s^{\perp}$  is slightly larger than $n_s^{\parallel}$
while the situation is inverted in the case of  Er BEC with $\epsilon_{\text{dd}}=0.38$ where  $n_s^{\parallel} > n_s^{\perp}$.
This reveals that QG effects on the superfluidity could be pronounced in perpendicular direction rather than in the parallel direction depending on the relative strength of the DDI.
Another important remark is that the superfluid fraction increases with the  GUP parameter in contrast to the condensed fraction.
For sufficiently large values of $\overline\alpha$, it diverges from that of the ordinary dipolar BEC in both components. 
Our results would provide unique new insight into the physics of superfluidity in neutron stars \cite{Baym} and thus will furnish a probe of the neutron star interior.


\section{Experimental Test of Quantum Gravity} \label{Exp}

In order to constrain parameters in the previous GUP proposal, it is necessary to relate it to two observables namely: the condensed fraction (\ref{confra}) 
and the superfluid fraction obtained from Eqs.~(\ref{supflui1}) and (\ref{supflui2}).
In our analysis, two kinds of dipolar atomic systems are being treated. The first consists of ${}^{52}$Cr atoms with $s$-wave scattering length $a=100\, a_0$ ($a_0$ is the Bohr radius) 
and  relative DDI strength $\epsilon_{\text{dd}}=0.16$ \cite{Pfau}.
Next we consider ${}^{168}$Er BEC with $a=175\,a_0$, and $\epsilon_{\text{dd}}=0.38$ \cite{erbium}.
The average density of both species is $n=5\times 10^{20}$m$^{-3}$.


\begin{table}[h!]
\begin{center}
\begin{tabular} { cccc ccc} 
 \hline \hline\\
                           & $\epsilon_{\text{dd}}$ & $T/T_c^0 $  & $n_c/n $  &  $\alpha_0$  &  $\beta_0$ \\ 
 \hline\\
 ${}^{52}$Cr       & 0.16 & 0.20 & 95\%  & 3.42$\times 10^{22}$  & 1.17$\times 10^{45}$ \\   
    & 0.16 & 0.95 & 15\%  & 1.03$\times 10^{25}$  & 1.05$\times 10^{50}$ \\ 
 \hline \\
 ${}^{168}$Er     &0.38 & 0.20  & 97\%   &  $2.60 \times 10^{22}$  & 6.70$\times 10^{44}$\\
 &0.38 & 0.95  & 35\%   &  $7.76 \times 10^{24}$  & 6.03$\times 10^{49}$\\
 \hline\hline
\end{tabular}
\end{center}
\caption{Typical values of $\alpha_0$ and $\beta_0$ extracted from the superfluid fraction of ${}^{52}$Cr atoms \cite{Pfau} and ${}^{168}$Er atoms \cite{erbium}. }
\label{table:1}
\end{table}

\begin{table}[h!]
\begin{center}
\begin{tabular} { cccc ccc} 
 \hline \hline\\
                       & $T/T_c^0 $  & $ n_s^{\parallel}/n$  &  $n_s^{\perp}/n$  & $\alpha_0$  &  $\beta_0$ \\ 
 \hline\\
 ${}^{52}$Cr      & 0.35 &  $95\%$  &  $95\%$ & 3.42$\times 10^{24}$  & 1.17$\times 10^{49}$ \\   
                        & 0.75 &  $35\%$  &  $45\%$ & 6.85$\times 10^{25}$  & 4.70$\times 10^{51}$ \\   
 \hline \\
 ${}^{168}$Er    & 0.35  & $ 98\%$   &  $ 98\% $  & $2.60 \times 10^{24}$  & 6.70$\times 10^{48}$\\
                        & 1  & $ 25\%$   &  $ 15\% $  & $5.17 \times 10^{25}$  & 2.70$\times 10^{51}$\\
 \hline\hline
\end{tabular}
\end{center}
\caption{Typical values of $\alpha_0$ and $\beta_0$ extracted from the superfluid fraction of ${}^{52}$Cr atoms \cite{Pfau} and ${}^{168}$Er atoms \cite{erbium}. }
\label{table:2}
\end{table}

Table \ref{table:1} shows that  at sufficiently low temperature,  $T/T_c^0  \simeq 0.2$, where the ground-state population $n_c/n$ is large, 
our model predicts for the GUP parameters $\alpha_0\sim 10^{22}$ and $\beta_0\sim 10^{44}$. 
Clearly, these values improve the bounds set by the model of an ideal Bose gas \cite{Das} and  by the model of BEC with a pure contact interaction \cite{Boudj}.
The reason is that $\beta_0$  strongly depends on $\epsilon_{\text{dd}}$. For instance, $\beta_0$ (${}^{168}$Er) is one order of magnitude higher than $\beta_0$ (${}^{52}$Cr). 

However, the bounds on QG parameters obtained from measuring the superfluid fraction in both parallel and perpendicular directions
are $\alpha_0\sim 10^{24}$ and $\beta_0\sim 10^{48}$ as shown in Table \ref{table:2}.
Although these bounds are better than the results obtained for weakly interacting Bose gases \cite{Boudj},  they are weaker than 
those set by high-energy physics \cite{Fab} and measurements of an ideal Bose gas \cite{Das}.
Therefore, they are somehow not interesting compared to previous ones.

\section{Conclusions} \label{concl} 

In this paper we studied effects of QG due to GUP on the ground-state properties of dipolar BECs under the aim to constrain the GUP parameters.
Using the HFB-LQGUP approach we calculate corrections to the condensed fraction, the critical temperature, the EoS, and the superfluid fraction.
We showed that the intriguing interplay of QG and DDIs may significantly affect these quantities.
We also discussed the possible experimental tests of our theoretical predictions.
Our theory predicted that better bounds require a strong relative DDI strength and a large condensed fraction (i.e. low temperatures).
Compared to bounds set from high-energy physics and other experiments and theories \cite{Fab}, our bounds on the GUP obtained from the condensed fraction are better
while those obtained from the superfluid fraction are worse.
Our findings can be readily probed in current experiments, and might bring us closer to understanding whether gravity can be reconciled with quantum mechanics.

\end{document}